\documentclass[%
 reprint,
 amsmath,amssymb,
 aps,
]{revtex4-2}

\usepackage{graphicx}
\usepackage{dcolumn}
\usepackage{bm}

\newcommand{\phibar}{\bar{\varphi}}
\newcommand{\deltaphi}{\Delta\varphi}

\begin{document}

\preprint{APS/123-QED}

\title{Probing the physical origins of 1/f noise by nonlinear transport}

\author{Nicolas Fontaine}
\author{Bertrand Reulet}%
 \email{bertrand.reulet@usherbrooke.ca}
\affiliation{%
Département de physique, Institut quantique, Université de Sherbrooke, Sherbrooke, Québec J1K 2R1, Canada
}%

\date{\today}

\begin{abstract}
We report measurements of low frequency noise in the nonlinear electron transport of tunnel junctions that are dc or ac biased. We show that not only the resistance but also the nonlinear coefficients exhibit $1/f$-like noise, and that correlations among them exist. Using the common theoretical model for electron transport in tunnel junctions, we relate our measurements to fluctuations of barrier heights. We show in particular that the barrier heights on either side of the junction have strong fluctuations that are almost perfectly anti-correlated.
\end{abstract}

\maketitle


\textit{Introduction---}
The presence of low frequency noise, also called flicker or $1/f$ noise, is ubiquitous \cite{Buckingham:1983,VanDerZiel:1986}. It has been reported in numerous materials and devices, and is understood as due to slow fluctuators which influence the device's resistance \cite{Dutta:1981,Rogers:1984}. These resistance fluctuations are characterized by their spectral density $S_R(f)$ which grows like $1/f^\alpha$ at low frequency with $\alpha$ taking values between 0.5 and 2. Experimentally, these fluctuations are usually observed by biasing the device with a dc current $I$ and measuring the voltage noise spectral density $S_V(f)=S_R(f)I^2$. Despite many years of effort the physical origin of these resistance fluctuations remains unclear in many devices. With the advent of quantum technologies using Josephson junctions, 1/f noise has received renewed attention and has been recognized as a factor limiting their performance \cite{balandin2024}. 

1/f noise is usually attributed to fluctuations of the linear resistance. However, low-frequency noise was also observed in resistors driven by an ac current \cite{montagnon1954, lorteije1971, sutcliffe1971, jones1975, jones1976,may1976}, and the explanations proposed at the time did not fully account for it \cite{valHelvoort1977}. Zhigal’skii later suggested that the nonlinear coefficients of the I–V characteristic also fluctuate, assuming their fluctuations to be uncorrelated \cite{zhigal1997}. In tunnel junctions, conductance models show that the same barrier parameters enter the linear, quadratic and cubic terms \cite{brinkman1970}, so correlations between the fluctuations of the coefficients are expected. Here we combine dc and ac drives to measure separately the fluctuations of the linear and nonlinear coefficients and their correlations in metallic tunnel junctions. We find that these correlations can be significant, and using a standard tunneling model, we trace them to fluctuations of the barrier parameters.

\textit{Principle of the measurement---}
We suppose that the time-dependent voltage across the junction is given by:
\begin{equation}
    V(t)= R_0(t)I(t) + R_1(t)I(t)^2 + R_2(t)I(t)^3 \label{eq:VI}
\end{equation}
Here $I(t)$ is a time-dependent applied current and $R_i(t)$ slowly randomly fluctuating coefficients. The goal here is to measure the fluctuations of these $R_i(t)$ coefficients individually. The "usual" $1/f$ noise corresponds to the fluctuations of $R_0(t)$. A dc excitation $I(t)=I_{0}$ gives rise to low frequency voltage noise composed of a mixture of spectral densities of $R_0$, $R_1$ and $R_2$ and their cross-correlations,
\begin{equation}
\begin{split}
    S_{DC}(f) = &S_R^{0,0}(f)I_0^2 + 2S_R^{0,1}(f)I_0^3 \\ &+ [2S_R^{0,2}(f)+S_R^{1,1}(f)] I_0^4 +...
\end{split}
\label{eq:mix}
\end{equation}
where $S_R^{i,j}$ is the spectral density of the correlations between $\delta R_i(t)$ and $\delta R_j(t)$: $S_R^{i,j}(f)=\mathrm{Re}\langle \delta R_i(f)\delta R_j(-f) \rangle$. It is interesting to separate the noise spectral density into its even and odd parts as a function of the dc current, effectively splitting Eq. (\ref{eq:mix}) into halves. By defining $S_\pm=(S_{DC}(I)\pm S_{DC}(-I))/2$, we find to leading order in $I_0$:
\begin{eqnarray}
    S_{+}(f) &=& S_R^{0,0}(f)I_0^2 + \mathcal{O}(I_0^4)\label{eq:S_symm}\\
    S_{-}(f) &=& 2S_R^{0,1}(f)I_0^3 + \mathcal{O}(I_0^5) \label{eq:S_asym}
\end{eqnarray}
In the following we consider not only the dc bias  case, but also an ac bias $I(t)=I_{0}\cos 2\pi f_0t$. Eq. (\ref{eq:VI}) supposes that the ac excitation frequency $f_0$ is not too high, so for example capacitive or inductive effects can be neglected. On the other hand, we want $f_0$ to be large enough so that it is beyond the frequency range where noise is measured. For the case of pure ac excitation, we find the noise spectral density at low frequency to be given by:
\begin{equation}
    S_{AC}(f) = \frac14 S_R^{1,1}(f) I_0^4 \label{eq:Sac}
\end{equation}
where the spectral density of $R_1$ is now isolated. Eq.(\ref{eq:Sac}) describes very well the results of a recent experiment \cite{fontaine2023} where the noise in the presence of an ac bias was shown to scale like $I_0^4$.

\textit{Experimental setup---}
\begin{figure}
    \centering
    \includegraphics[width=0.95\linewidth]{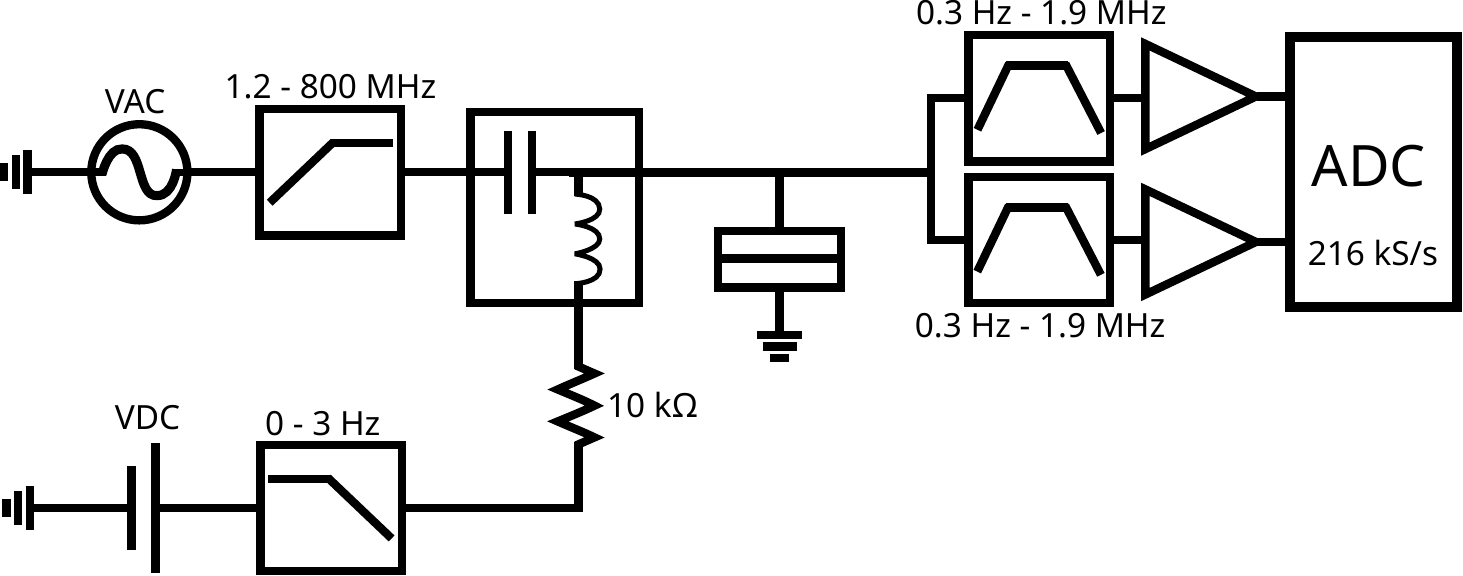}
    \caption{Experimental setup for the measurement of the low frequency voltage noise $S_V(f)$ with ac and/or dc excitation.}
    \label{fig:montageDC}
\end{figure}
We worked with two tunnel junctions of resistance close to $50\;\Omega$. An Al/AlOx/Al junction of area $4\;\mu\mathrm{m}^2$ fabricated by usual photolithography and metal deposition \cite{Spietz2003}, and a Nb/AlOx/Nb junction of area $78.5\;\mu\mathrm{m}^2$ manufactured by StarCryogenics. The setup for the noise measurement at frequencies up to 100 kHz is identical to that of \cite{fontaine2023}, and is shown in Fig. \ref{fig:montageDC}: the sample, mounted in a closed copper box and connected to an SMA connector, is current biased through the dc port of a bias tee. The voltage across the sample is amplified by two voltage amplifiers in parallel, then digitized. The voltage noise spectral density is obtained by cross-correlation of the two voltage measurements, which renders the contribution of the amplifier noise to the measured noise negligible. The setup is sensitive enough to measure thermal noise of a 50$\;\Omega$ resistor at room temperature, which corresponds to $4k_BTR=0.8\;\mathrm{(nV)}^2/\mathrm{Hz}$ with $k_B$ the Boltzmann constant and $T$ the temperature. The ac excitation is performed through the RF port of the bias tee. We have chosen to use an excitation of frequency $f_0=7.7$ MHz, much higher than the bandwidth of the $1/f$ noise, and low enough to avoid problems with attenuation and propagation delay in the cables. No $1/f$ noise was visible when replacing the junction by a $50\;\Omega$ metallic resistor. Thus neither the DC source nor the AC one generated low frequency noise.

\begin{figure}
    \centering
    \includegraphics[width=\linewidth]{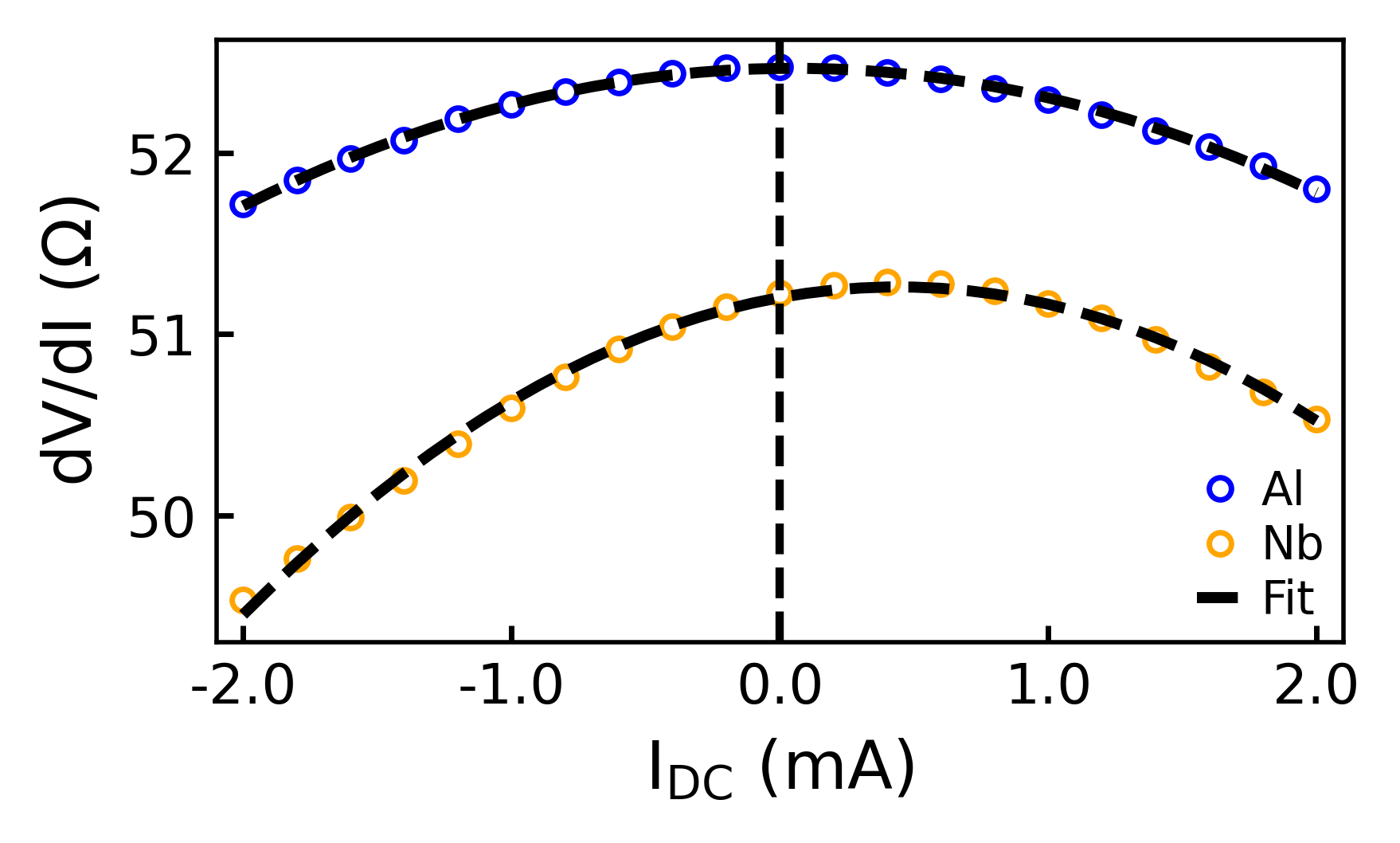}
    \caption{Differential resistance of the junctions as a function of dc bias current. Blue and orange symbols are measurements for the Al and Nb junctions respectively. Black dashed lines are parabolic fits from which we deduce $R_0=52.47\;\Omega$, $R_1=9.19\;\Omega.\mathrm{A}^{-1}$ and $R_2=-60190\;\Omega.\mathrm{A}^{-2}$ for the Al junction and $R_0=51.20\;\Omega$, $R_1=133.67\;\Omega.\mathrm{A}^{-1}$ and $R_2=-101197\;\Omega.\mathrm{A}^{-2}$ for the Nb junction.}
    \label{fig:dVdI}
\end{figure}

\textit{Measurements \& Results---}
The differential resistance $dV/dI$ of the junctions is shown in Fig. \ref{fig:dVdI} as a function of the applied DC current. For both junctions it follows a parabolic behaviour, as commonly observed and modeled for metal-based junctions \cite{simmons1963,brinkman1970}. For the Al junction, the measured nonlinearity appears weak, the differential resistance falling by only $1.5\%$ at $I=2$ mA from its value at zero bias. $dV/dI$ is an even function of the bias current (a parabolic fit of the data gives the maximum resistance occurring at $\sim50\;\mu$A). The Nb junction shows a somewhat stronger nonlinearity but more importantly a much stronger asymmetry: the maximum of $dV/dI$ occurs at $I=0.5$ mA. These characteristics translate into $R_1$ almost vanishing for the Al junction and $R_2$ being larger for the Nb junction than for the Al one. The reason for the different behaviour between the two junctions is probably due to the fabrication process: the Al junction is obtained by oxidization of an Al film followed by Al deposition, so the junction is symmetric. In contrast, the Nb junction is obtained by deposition of a Nb/Al bilayer followed by oxidization and deposition of a Nb film on the top. Such a junction is clearly not symmetric. If the oxidization process does not fully reach the bottom of the Al film, a Nb/Al interface remains. The process could also reach and damage the Nb layer, modifying it. The former is most probable since these junctions are meant to be used as Josephson junctions at low temperature.

\begin{figure}
    \centering
    \includegraphics[width=\linewidth]{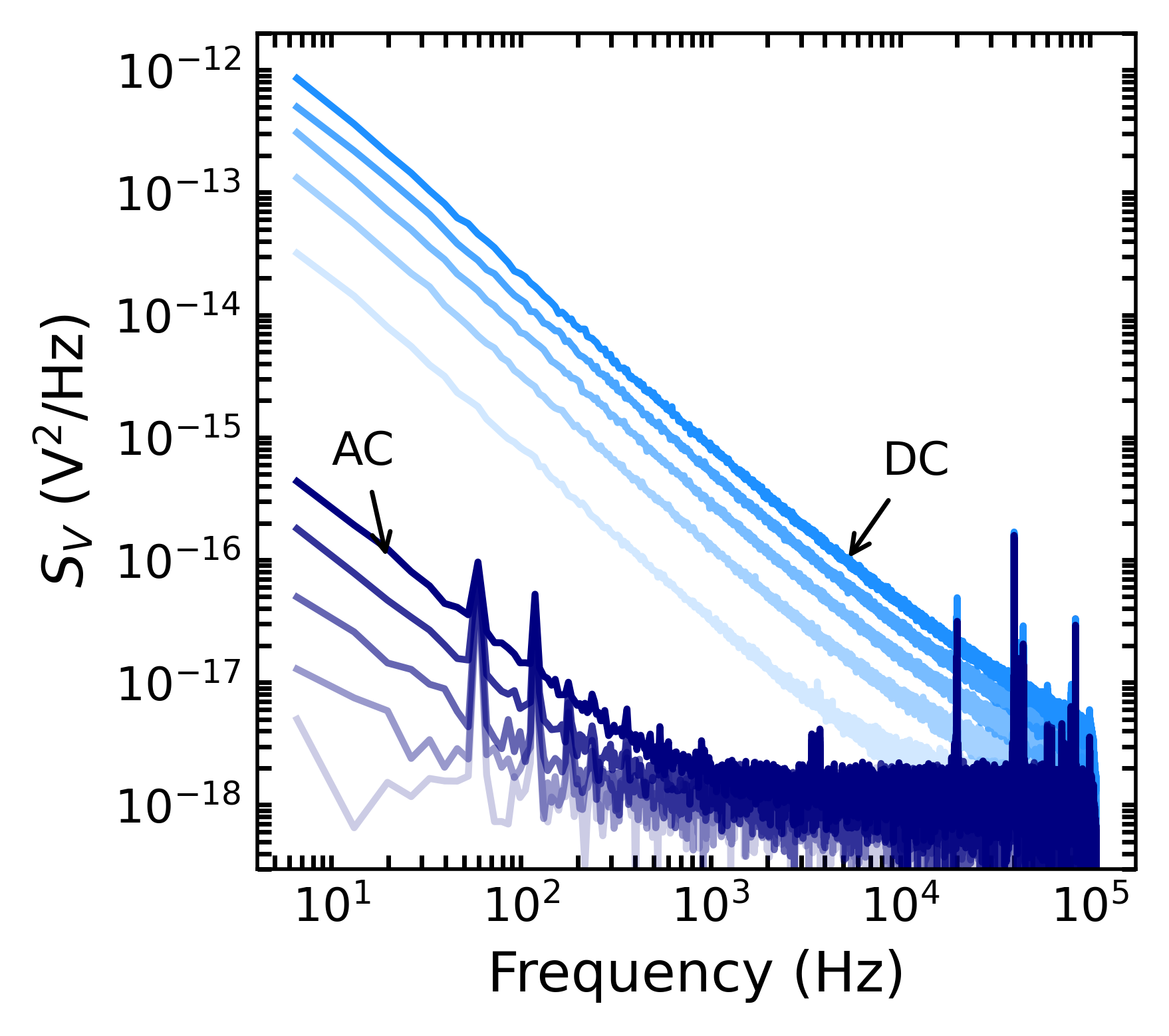}
    \caption{Raw noise spectra of the Al junction for ac (navy) and positive dc (light blue) biases regularly spaced  between 0.4 and 2.0 mA.}
    \label{fig:spectresRaw}
\end{figure}

Raw noise spectra of the Al junction are shown on Fig. \ref{fig:spectresRaw} for ac (navy) and positive dc (light blue) biases regularly spaced  between 0.4 and 2.0 mA. We see that the shape of the spectra does not change under stronger biases, the 1/f noise simply rises with the bias, in agreement with Eqs.(\ref{eq:S_symm},\ref{eq:Sac}). We measure the same behaviour with the Nb junction (see Sec.\ref{sec:raw} of Supplemental Material \cite{supmat}). More precisely, we  confirm that  $S_+\propto I_0^2$, $S_-\propto I_0^3$ (for the Nb junction, $S_-$ vanishes within uncertainty for the Al one), and $S_{AC}\propto I_0^4$, see section \ref{sec:NoiseVsBias} of Supplemental Material \cite{supmat}. 

Our results are fully consistent with the hypothesis of fluctuating nonlinear resistance coefficients described by Eq.(\ref{eq:VI}) . From the measurements of $S_\pm$ and $S_{AC}$ we extract $S_R^{0,0}$, $S_R^{1,1}$ and $S_R^{0,1}$: $S_+$ provides the amplitude of the fluctuations of $R_0$, $S_{AC}$ that of $R_1$ and $S_-$ the correlations between $R_0$ and $R_1$. See more in section \ref{sec:extractSR} of Supplemental Material \cite{supmat}.

\begin{figure}
    \centering
    \includegraphics[width=\linewidth]{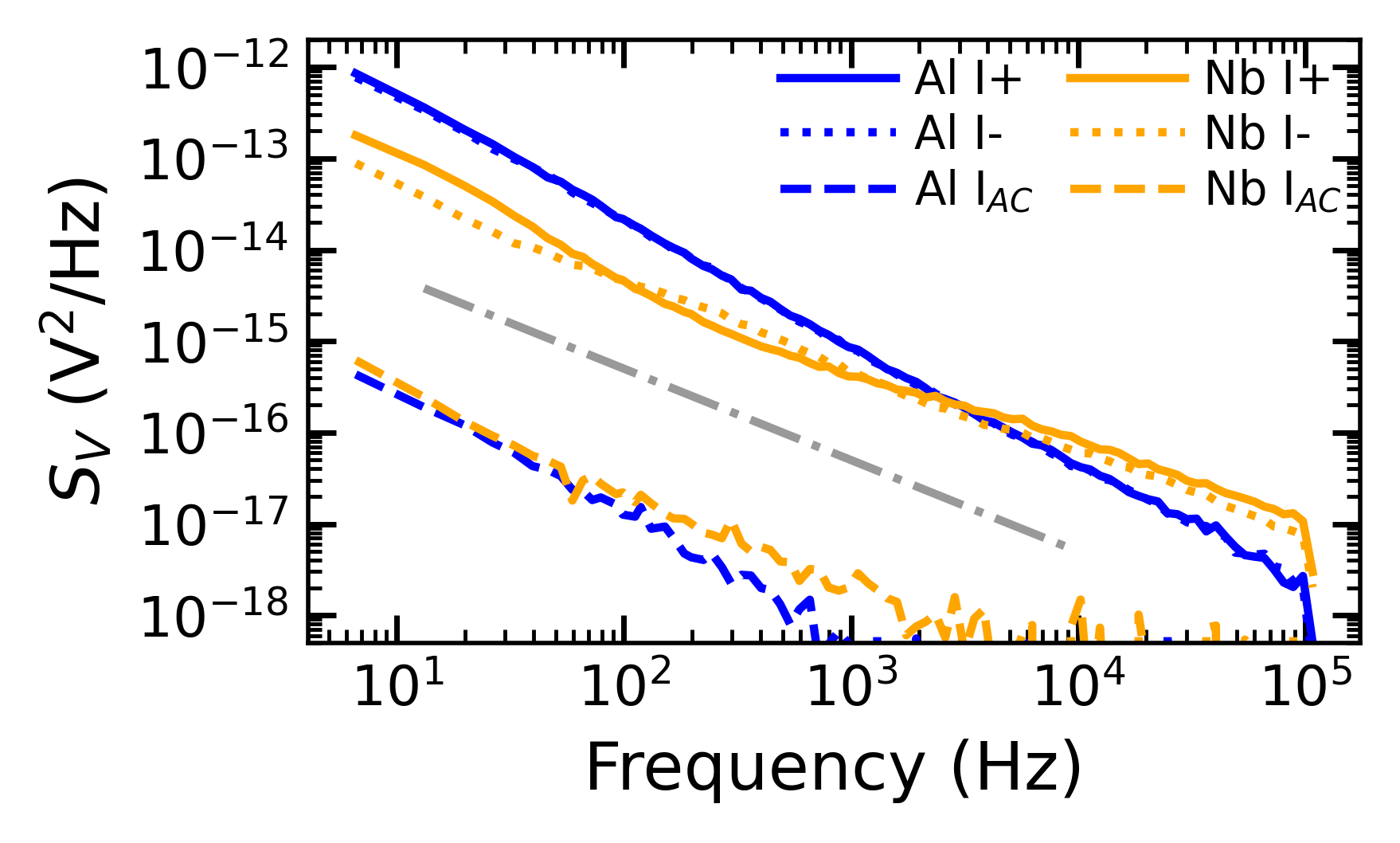}
    \caption{Noise spectra of the aluminum junction (blue) and niobium junction (orange) for positive dc (solid), negative dc (dotted) and ac (dashed) biases. Thermal noise, shot noise and residual setup noise were removed. Data points were downsampled for visibility. The gray dash-dotted line provides a guide to the eye of $1/f^\alpha$ with $\alpha=1$.}
    \label{fig:spectres}
\end{figure}

The noise spectra of the two junctions at maximum bias are presented in Fig. \ref{fig:spectres} in log-log scale. To obtain those spectra, the zero bias spectrum, which includes thermal noise and the setup noise floor of $\sim9\times10^{-19}$ V$^2$/Hz, was subtracted from all spectra. The shot noise of the junction was also removed ($9\times10^{-19}$ V$^2$/Hz for dc and $5\times10^{-19}$ V$^2$/Hz for ac biases)\cite{Spietz2003}. Raw measurements are available in Sec. \ref{sec:raw} of Supplemental Material \cite{supmat}. These corrections mostly affect the high frequency part of the spectra ($>$ 10 kHz). Only the residual 1/f noise of the junction is shown. For the Al junction (blue), positive (solid line) and negative (dotted line) dc biases lead to the same noise spectra, which appear linear with a slope of $1/f^{1.35}$. The Nb junction (orange) produces notably different spectra under positive and negative dc biases. Both spectra present smooth bumps and dips, but at different frequencies. The noise level of both spectra falls roughly as $1/f^{0.95}$ though it is harder to estimate given their wavy aspect. Under an ac bias (dashed lines), both junctions produce similar noise level and shape. Noise measured with an ac bias is several orders of magnitude smaller than under a dc bias \cite{jones1975,fontaine2023}.

\textit{Discussion---}
We now relate the measured nonlinear transport coefficients and their fluctuations to physical parameters of the tunnel junction. Conductance in tunnel junctions is usually calculated from the transmission through a potential barrier \cite{harrison1961}. The Wentzel–Kramers–Brillouin (WKB) approximation made it possible to get analytical expressions first for symmetrical junctions \cite{simmons1963,simmons1964}, later extended to asymmetrical ones \cite{brinkman1970}. Although recent developments made the symmetrical conductance models more precise \cite{raisanen2026}, we will use the Brinkman-Dynes-Rowell (BDR) model \cite{brinkman1970} which is still the simplest that retains asymmetry, and is widely used to describe transport in tunnel junctions. Modeling the junction by a trapezoidal barrier of thickness $d$ , and heights $q\varphi_{1,2}$, the conductance $G(V)=dI(V)/dV$ is given by:
\begin{align}
    \frac{G(V)}{G_0} = 1 - \frac{\sqrt{2m_eq}}{12\hbar}\frac{d\deltaphi}{\phibar^{3/2}}V + \frac{m_eq}{4\hbar^2}\frac{d^2}{\phibar}V^2
    \label{eq:brinkman}
\end{align}
with $\phibar = (\varphi_1+\varphi_2)/2$, $\deltaphi = \varphi_2-\varphi_1$, $q$ the elementary charge and $m_e$ the electron mass. $q\phibar$ is the average barrier height and $q\deltaphi$ its asymmetry. The conductance at zero bias $G_0$ takes the form $G_0 = S\frac{q^2\sqrt{2m_eq}}{4\pi^2\hbar^2}\frac{\sqrt{\phibar}}{d}\exp\left(-\frac{2\sqrt{2m_eq}}{\hbar}d\sqrt{\phibar}\right)$. $S$ is the area of the tunnel junction.

The constant, linear and quadratic terms of Eq. (\ref{eq:brinkman}) are simply related to the coefficients $R_0$, $R_1$ and $R_2$ of the junction (see expressions in Sec. \ref{sec:BDRframework} of Supplemental Material \cite{supmat}).  From our measurements of $R_{0,1,2}$ we find $d=7.50$ \AA, $\phibar=1.48$ V and $\deltaphi=0.037$ V for the Al junction and $d=9.85$ \AA, $\phibar=1.40$ V and $\deltaphi=0.40$ V for the Nb one. These parameters are in line with those in \cite{Dorneles2003,Gloos2003,Cyster2020} for Al junctions with a resistance-area product $R_0S\sim220\;\Omega\mu \mathrm{m}^2$ and those presented in \cite{Kang2014,Paramonov2019,Paramonov2020} for Nb junctions with $R_0S\sim4000\;\Omega\mu \mathrm{m}^2$. The Al junction has a much smaller area, yet the same resistance as the Nb one because its oxide barrier is thinner. As a consequence, it is also more linear (its $R_2$ is smaller). The asymmetry $\deltaphi$ is also vastly different in both junctions, as was suggested by their differential resistance. 

For the remainder of the article, we will discuss the fluctuations of $R_0$, $R_1$ and their correlations, whose contributions dominate at not too large bias current. Within the theoretical model, the fluctuations of the coefficients $R_{0,1}$ are related to those of $d$, $\phibar$ and $\deltaphi$ and the correlations among them. While $\deltaphi$ plays a special role, $d$ and $\phibar$ appear mainly together through $d\sqrt{\phibar}$, so we cannot distinguish their fluctuations. To simplify the discussion we will suppose that the thickness $d$ is constant while the barrier height and asymmetry fluctuate (we discuss fluctuations of the barrier thickness in Sec.\ref{sec:dnotzero} of Supplemental Material \cite{supmat}). Their spectral densities are related to those of $R_i$ by:
\begin{eqnarray}
S_{\phibar\phibar} &=&  \frac{S_R^{0,0}}{\langle R_0\rangle^2k_2^2} \label{eq:Sphi}\\
S_{\phibar\deltaphi} &=&  \frac{\langle\deltaphi\rangle}{k_2}\left(\frac{S_R^{0,1}}{\langle R_0\rangle \langle R_1\rangle} - k_2k_4 S_{\phibar\phibar}\right) \label{eq:Sphidelta}\\
S_{\deltaphi\deltaphi} &=& \langle\deltaphi\rangle^2\left(\frac{S_R^{1,1}}{\langle R_1\rangle^2} - k_4^2S_{\phibar\phibar} - \frac{2k_4}{\langle\deltaphi\rangle} S_{\phibar\deltaphi}\right) \label{eq:Sdelta}
\end{eqnarray}
where $k_2 = \frac{\eta-1}{2\langle\phibar\rangle}$, $k_4 = \frac{2\eta-5}{2\langle\phibar\rangle}$ and $\eta = \frac{2\sqrt{2m_eq}}{\hbar} d\sqrt{\langle\phibar\rangle}$.

\begin{figure}
    \centering
    \includegraphics[width=\linewidth]{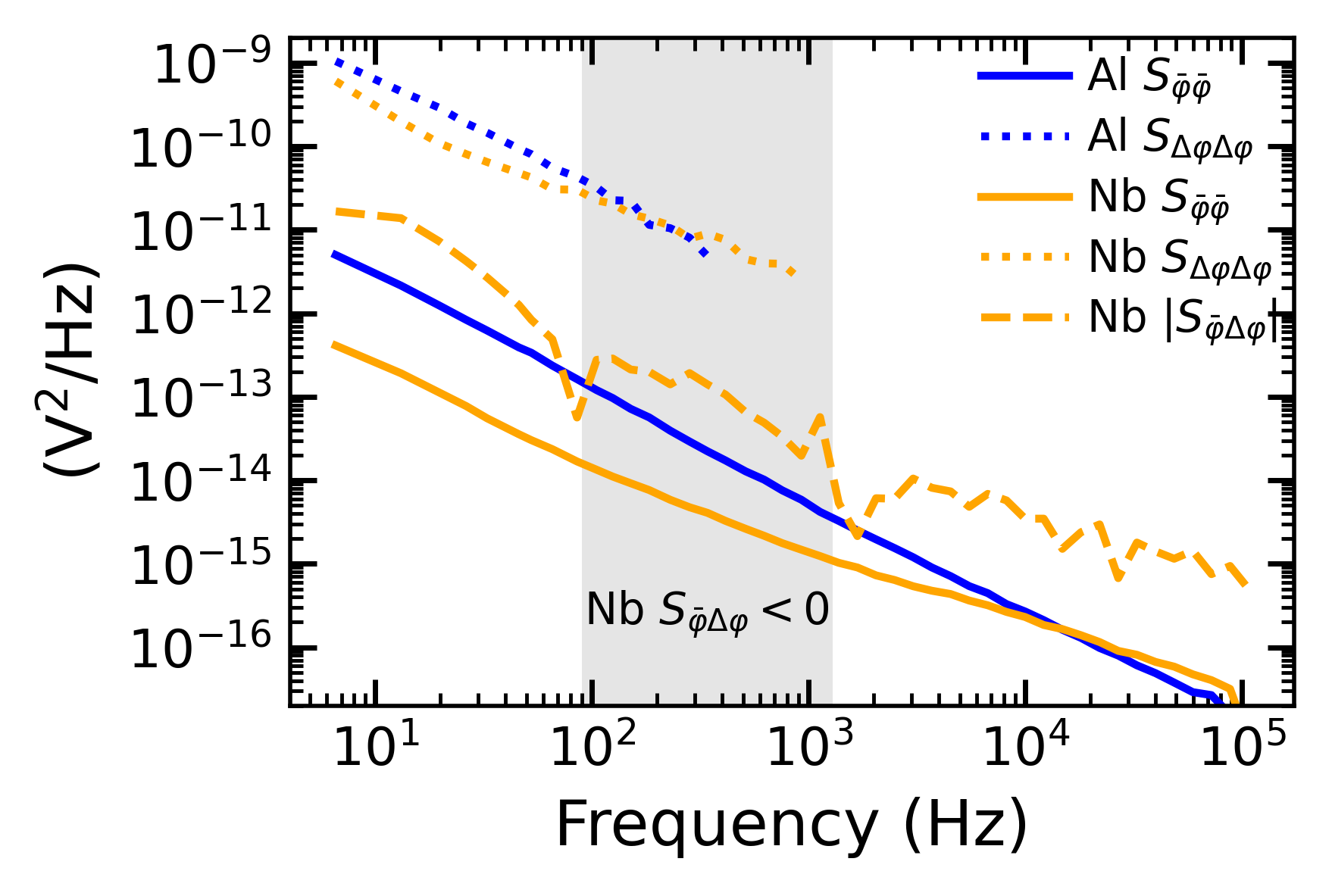}
    \caption{$S_{\deltaphi\deltaphi}$ (dotted lines), $S_{\phibar\phibar}$ (solid lines) and $S_{\phibar\deltaphi}$ (dashed line) for the Al (blue) and Nb (orange) junctions. $S_{\phibar\deltaphi}$ is very small for the Al junction and is not plotted, and it takes negative values for the Nb junction in the shaded area. Data points have been downsampled for visibility.}
    \label{fig:psdBrinkman}
\end{figure}

The resulting $S_{\phibar\phibar}$, $S_{\phibar\deltaphi}$ and $S_{\deltaphi\deltaphi}$ are shown in Fig. \ref{fig:psdBrinkman}. The measured $S_{\deltaphi\deltaphi}$ (dotted lines), i.e., the variances of $\deltaphi$, appear similar for the Al (blue) and Nb (orange) junctions, although the average asymmetry $\langle\deltaphi\rangle$ for the Al junction almost vanishes. Their measured levels are orders of magnitude greater than the other spectral densities. $S_{\phibar\phibar}$ (solid line), i.e. the variances of $\phibar$ are similar for both junctions although, as seen previously, their 1/f exponent $\alpha$ differ. A crucial difference between the junctions appears in $S_{\phibar\deltaphi}$ (dashed line), which measures the correlation between the fluctuations of $\phibar$ and those of $\deltaphi$. It is below our measurement resolution for the Al junction and clearly present for the Nb junction. For the latter, it changes sign as a function of frequency, being negative, which corresponds to anti-correlations,  in the gray zone from 90 Hz to 1300 Hz.

For a symmetric junction ($\langle\deltaphi\rangle\to 0$), only the first terms of Eqs. (\ref{eq:Sphidelta},\ref{eq:Sdelta}) remain since $\langle R_1\rangle \propto \langle\deltaphi\rangle$. In that case, $S_{\phibar\deltaphi} \propto S_R^{0,1}(f)$ and $S_{\deltaphi\deltaphi} \propto S_R^{1,1}(f)$: the noise of a symmetric junction provides direct measurements of the spectral densities of the fluctuations of $\deltaphi$ and their correlations with the fluctuations of $\phibar$. The experiments on the Al junction thus demonstrate that even though $\deltaphi$ is almost zero on average, its fluctuations are important and appear not correlated with those of $\phibar$ within our measurement resolution.  For an asymmetric junction, $\phibar$ and $\deltaphi$ are not independent: they can be correlated or anti-correlated, depending on frequency. However for any junction, even though fluctuations of $\deltaphi$ are orders of magnitude larger than those of $\phibar$, they do not influence the "usual" $1/f$ noise $S_R^{0,0}$, which is mostly due to the fluctuations of $\phibar$ (or $d$), see Eq.(\ref{eq:Sphi}).

It is insightful to rephrase these conclusions in terms of the fluctuations of the barrier heights on each side of the junction $\varphi_{1,2}$: $S_{\deltaphi\deltaphi}$ being the dominant fluctuation implies that $\varphi_1$ and $\varphi_2$ must be strongly anti-correlated. Indeed, we find that the correlation coefficient $S_{\varphi_1\varphi_2}/\sqrt{ S_{\varphi_1\varphi_1}S_{\varphi_2\varphi_2} }$ equals -0.96 for the Al junction and -1 for the Nb one, both are frequency independent, see Fig. \ref{fig:correlationsPhis} of Supplemental Material \cite{supmat}. Moreover, for the Al junction, $S_{\varphi_1\varphi_1}\simeq S_{\varphi_2\varphi_2}$ suggests that the two metal/insulator interfaces have similar quality, contrary to the Nb junction. (more details in Sec.\ref{sec:implications} of Supplemental Material \cite{supmat}).

\textit{Conclusion---}
We have used the effect of ac and positive/negative dc bias on the low frequency noise of nonlinear metal tunnel junctions to elucidate which physical parameters cause the fluctuations of the nonlinear transport coefficients. We find that barrier heights on both sides of the junction strongly fluctuate, dominating the noise of nonlinear transport. However, they are almost perfectly anti-correlated so they do not affect the "usual" $1/f$ noise of linear transport. This strongly calls for new microscopic theoretical models. For example, the presence of charged traps/defects on each interface could be almost balanced by their image charge on the other interface, thanks to the strong screening in the metallic contacts. This would result in the barrier asymmetry $\deltaphi$ fluctuating much more than its height $\phibar$, as we observe. If this is correct, a weaker anti-correlation should be observed with semiconducting contacts. Moreover our experimental approach can be applied to many other conducting samples to better understand the physical origins of their flicker noise.

\textit{Acknowledgments---}
We thank A. Dumont, G. Laliberté and C. Lupien for their help with the experimental techniques and E. Pinsolle for providing us with the Al junction. We are grateful to A. Bouhelier and A.-M. Tremblay for fruitful discussions. We thank Quantum e-Motion for providing us with Nb samples and for financial support. This work was supported by the Canada Research Chair program, the NSERC, the Canada First Research Excellence Fund, the FRQNT, and the Canada Foundation for Innovation.

\bibliography{biblio}

@book{vanDerZiel:1986,
  title={Noise in Solid State Devices and Circuits},
  author={Van der Ziel, A.},
  isbn={9780471832348},
  lccn={lc85029572},
  url={https://books.google.ca/books?id=RjJTAAAAMAAJ},
  year={1986},
  publisher={Wiley}
}

@book{buckingham:1983,
  title={Noise in Electronic Devices and Systems},
  author={Buckingham, M.J.},
  isbn={9780853122180},
  lccn={lc83008605},
  series={Electrical and Electronic Engineering Series},
  url={https://books.google.ca/books?id=xP1SAAAAMAAJ},
  year={1983},
  publisher={E. Horwood}
}

@article{dutta:1981,
  title = {Low-frequency fluctuations in solids: $\frac{1}{f}$ noise},
  author = {Dutta, P. and Horn, P. M.},
  journal = {Rev. Mod. Phys.},
  volume = {53},
  issue = {3},
  pages = {497--516},
  numpages = {0},
  year = {1981},
  month = {Jul},
  publisher = {American Physical Society},
  doi = {10.1103/RevModPhys.53.497},
  url = {https://link.aps.org/doi/10.1103/RevModPhys.53.497}
}

@article{rogers:1984,
  title = {Composition of $\frac{1}{f}$ Noise in Metal-Insulator-Metal Tunnel Junctions},
  author = {Rogers, C. T. and Buhrman, R. A.},
  journal = {Phys. Rev. Lett.},
  volume = {53},
  issue = {13},
  pages = {1272--1275},
  numpages = {0},
  year = {1984},
  month = {Sep},
  publisher = {American Physical Society},
  doi = {10.1103/PhysRevLett.53.1272},
  url = {https://link.aps.org/doi/10.1103/PhysRevLett.53.1272}
}

@article{balandin2024,
    author = {Balandin, Alexander A. and Paladino, Elisabetta and Hakonen, Pertti J.},
    title = {Electronic noise—From advanced materials to quantum technologies},
    journal = {Applied Physics Letters},
    volume = {124},
    number = {5},
    pages = {050401},
    year = {2024},
    month = {01},
    issn = {0003-6951},
    doi = {10.1063/5.0197142},
    url = {https://doi.org/10.1063/5.0197142}
}

@INPROCEEDINGS{fontaine2023,
  author={Fontaine, Nicolas and Dumont, Alexandre and Reulet, Bertrand},
  booktitle={2023 International Conference on Noise and Fluctuations (ICNF)}, 
  title={Low frequency noise in AC biased metallic tunnel junctions}, 
  year={2023},
  volume={},
  number={},
  pages={1-4},
  doi={10.1109/ICNF57520.2023.10472762}}

@article{montagnon1954,
  title={L.F. noise in resistors excited by r.f. currents},
  author={Montagnon, N.B.},
  journal={Wireless Eng.},
  volume={31},
  pages={255--265,301--305},
  year={1954},
}

@article{lorteije1971,
  title={AMPLITUDE MODULATION BY 1/F NOISE IN RESISTORS RESULTS IN 1/DELTAF NOISE},
  author={Lorteije, JH and Hoppenbrouwers, AM},
  journal={Philips Research Reports},
  volume={26},
  number={1},
  pages={29--+},
  year={1971},
  publisher={PHILIPS RESEARCH LABORATORIES BUILDING WBP ROOM NO. 42, EINDHOVEN, NETHERLANDS}
}

@inproceedings{sutcliffe1971,
  title={Measurement of fluctuations of conductance with DC and AC excitation},
  author={Sutcliffe, H},
  booktitle={Noise Conference Proceedings: Le Bruit de Fond des Composants Actifs Semi-Conducteurs},
  pages={39--42},
  year={1971}
}

@article{jones1975,
  title={Direct correlation between 1/f and other noise sources},
  author={Jones, BK and Francis, JD},
  journal={Journal of Physics D: Applied Physics},
  volume={8},
  number={10},
  pages={1172--1176},
  year={1975}
}

@article{jones1976,
  title={1/f and 1/$\Delta$f noise produced by a radio-frequency current in a carbon resistor},
  author={Jones, BK},
  journal={Electronics Letters},
  volume={12},
  number={4},
  pages={110--111},
  year={1976},
  publisher={IET}
}

@article{may1976,
  title={1/f noise in multicarrier systems},
  author={May, EJP and Morgan, HG},
  journal={Electronics Letters},
  volume={12},
  number={1},
  pages={8--9},
  year={1976},
  publisher={IET}
}

@article{valHelvoort1977,
author = {G.J.M. van Helvoort  and H.G.E. Beck },
title = {Model for the excitation of 1/f noise by high-frequency a.c. signals},
journal = {Electronics Letters},
volume = {13},
issue = {18},
pages = {542-544},
year = {1977},
doi = {10.1049/el:19770390},
URL = {https://digital-library.theiet.org/doi/abs/10.1049/el%3A19770390}
}

@article{zhigal1997,
  title={1/f noise and nonlinear effects in thin metal films},
  author={Zhigalskii, Gennadii P},
  journal={Physics-uspekhi},
  volume={40},
  number={6},
  pages={599--622},
  year={1997}
}

@article{brinkman1970,
    author = {Brinkman, W. F. and Dynes, R. C. and Rowell, J. M.},
    title = {Tunneling Conductance of Asymmetrical Barriers},
    journal = {Journal of Applied Physics},
    volume = {41},
    number = {5},
    pages = {1915-1921},
    year = {1970},
    month = {04},
    issn = {0021-8979},
    doi = {10.1063/1.1659141},
    url = {https://doi.org/10.1063/1.1659141}
}

@article{harrison1961,
  title = {Tunneling from an Independent-Particle Point of View},
  author = {Harrison, Walter A.},
  journal = {Phys. Rev.},
  volume = {123},
  issue = {1},
  pages = {85--89},
  numpages = {0},
  year = {1961},
  month = {Jul},
  publisher = {American Physical Society},
  doi = {10.1103/PhysRev.123.85},
  url = {https://link.aps.org/doi/10.1103/PhysRev.123.85}
}

@article{simmons1963,
    author = {Simmons, John G.},
    title = {Generalized Formula for the Electric Tunnel Effect between Similar Electrodes Separated by a Thin Insulating Film},
    journal = {Journal of Applied Physics},
    volume = {34},
    number = {6},
    pages = {1793-1803},
    year = {1963},
    month = {06},
    issn = {0021-8979},
    doi = {10.1063/1.1702682},
    url = {https://doi.org/10.1063/1.1702682}
}

@article{simmons1964,
    author = {Simmons, John G.},
    title = {Generalized Thermal J‐V Characteristic for the Electric Tunnel Effect},
    journal = {Journal of Applied Physics},
    volume = {35},
    number = {9},
    pages = {2655-2658},
    year = {1964},
    month = {09},
    issn = {0021-8979},
    doi = {10.1063/1.1713820},
    url = {https://doi.org/10.1063/1.1713820}
}

@article{raisanen2026,
    author = {Räisänen, Ilmo M. W. and Maasilta, Ilari J.},
    title = {Improvement of the Simmons model for tunnel junctions},
    journal = {Journal of Applied Physics},
    volume = {139},
    number = {7},
    pages = {073902},
    year = {2026},
    month = {02},
    issn = {0021-8979},
    doi = {10.1063/5.0304889},
    url = {https://doi.org/10.1063/5.0304889}
}

@article{dorneles2003,
    author = {Dorneles, L. S. and Schaefer, D. M. and Carara, M. and Schelp, L. F.},
    title = {The use of Simmons’ equation to quantify the insulating barrier parameters in Al/AlOx/Al tunnel junctions},
    journal = {Applied Physics Letters},
    volume = {82},
    number = {17},
    pages = {2832-2834},
    year = {2003},
    month = {04},
    issn = {0003-6951},
    doi = {10.1063/1.1569986},
    url = {https://doi.org/10.1063/1.1569986}
}

@article{gloos2003,
doi = {10.1088/0953-8984/15/10/320},
url = {https://doi.org/10.1088/0953-8984/15/10/320},
year = {2003},
month = {mar},
publisher = {},
volume = {15},
number = {10},
pages = {1733},
author = {K Gloos and P J Koppinen and J P Pekola},
title = {Properties of native ultrathin aluminium oxide tunnel barriers},
journal = {Journal of Physics: Condensed Matter}
}

@article{cyster2020,
  title = {Effect of atomic structure on the electrical response of aluminum oxide tunnel junctions},
  author = {Cyster, M. J. and Smith, J. S. and Vaitkus, J. A. and Vogt, N. and Russo, S. P. and Cole, J. H.},
  journal = {Phys. Rev. Res.},
  volume = {2},
  issue = {1},
  pages = {013110},
  numpages = {10},
  year = {2020},
  month = {Jan},
  publisher = {American Physical Society},
  doi = {10.1103/PhysRevResearch.2.013110},
  url = {https://link.aps.org/doi/10.1103/PhysRevResearch.2.013110}
}

@article{kang2014,
title = {Measurements of tunneling barrier thicknesses for Nb/Al–AlOx/Nb tunnel junctions},
journal = {Physica C: Superconductivity and its Applications},
volume = {503},
pages = {29-32},
year = {2014},
issn = {0921-4534},
doi = {https://doi.org/10.1016/j.physc.2014.05.006},
url = {https://www.sciencedirect.com/science/article/pii/S0921453414001683},
author = {Xinjie Kang and Liliang Ying and Hai Wang and Guofeng Zhang and Wei Peng and Xiangyan Kong and Xiaoming Xie and Zhen Wang}
}

@article{paramonov2019,
  title={Determination of the Parameters of Tunneling Barriers of Superconducting Tunnel Structures for Submillimeter Receivers},
  author={Maxim E. Paramonov and Lyudmila V. Filippenko and Pavel N. Dmitriev and Mikhail Fominsky and V. P. Koshelets},
  journal={Journal of Communications Technology and Electronics},
  year={2019},
  volume={64},
  pages={1144 - 1148},
  url={https://api.semanticscholar.org/CorpusID:209940801}
}

@article{paramonov2020,
author={Paramonov, M. E.
and Filippenko, L. V.
and Dmitriev, P. N.
and Fominsky, M. Yu.
and Ermakov, A. B.
and Koshelets, V. P.},
title={Parameters of the Tunnel Barrier of Superconducting Niobium-Based Structures},
journal={Physics of the Solid State},
year={2020},
month={Sep},
day={01},
volume={62},
number={9},
pages={1534-1538},
issn={1090-6460},
doi={10.1134/S1063783420090231},
url={https://doi.org/10.1134/S1063783420090231}
}

@misc{supmat,
  note = {See Supplemental Material at [link provided by publisher] for raw measurements of $S_{\pm}$ and $S_{AC}$ and their response to bias, methodology for the extraction of $S_R^{x,y}$ spectra, expressions of $R_n$ coefficients in terms of BDR parameters, a discussion on the implications of barrier thickness fluctuations, a discussion on junction interface quality and the implications of our results on barrier parameter fluctuations.}
}

@article{Spietz2003,
author = {Lafe Spietz  and K. W. Lehnert  and I. Siddiqi  and R. J. Schoelkopf },
title = {Primary Electronic Thermometry Using the Shot Noise of a Tunnel Junction},
journal = {Science},
volume = {300},
number = {5627},
pages = {1929-1932},
year = {2003},
doi = {10.1126/science.1084647}
}

\newpage

\section{Supplemental material}

\subsection{Raw noise spectra}
\label{sec:raw}

Noise measurements of the Al junction with positive (a) and negative (b) dc, as well as ac (c) biases are shown in Fig. \ref{fig:spectresRawEd}. The noise level rises with the applied current, as expected, and the shape of the spectra stays identical. Positive and negative dc biases lead to the same spectra in shape and noise level.
\begin{figure*}
    \centering
    \includegraphics[width=\linewidth]{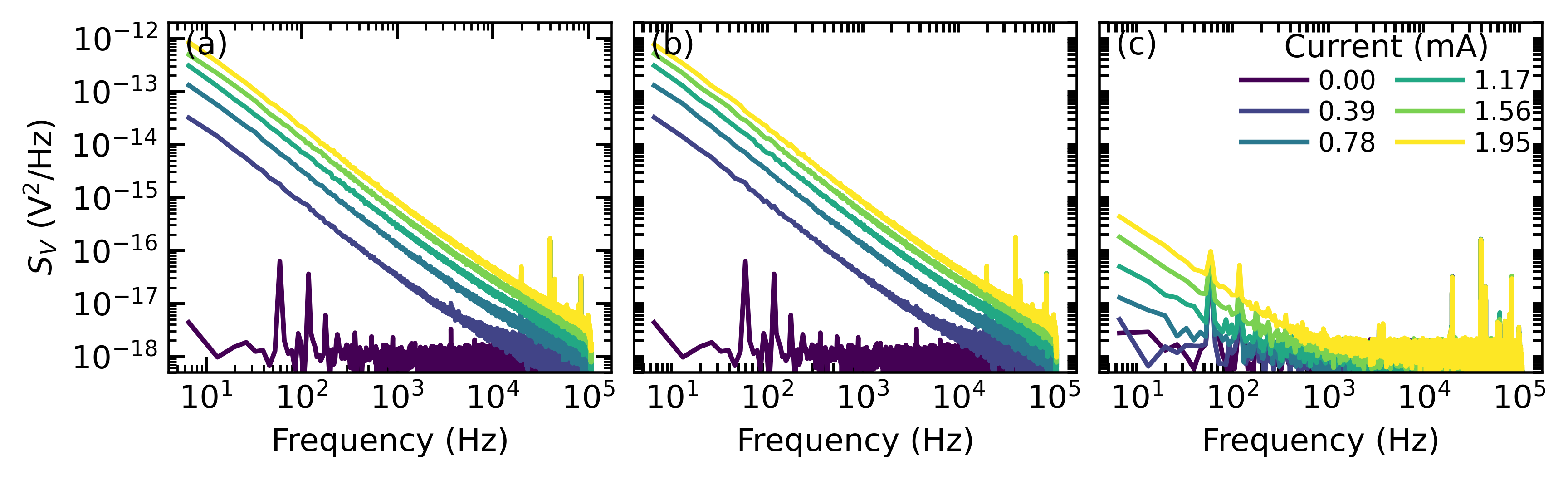}
    \caption{Raw noise measurements of the Al junction with positive (a) and negative (b) dc, as well as ac (c) biases.}
    \label{fig:spectresRawEd}
\end{figure*}

Noise measurements of the Nb junction with positive (a) and negative (b) dc, as well as ac (c) biases are shown in Fig. \ref{fig:spectresRawDec}. Here positive and negative dc biases lead to measurably different spectra.
\begin{figure*}
    \centering
    \includegraphics[width=\linewidth]{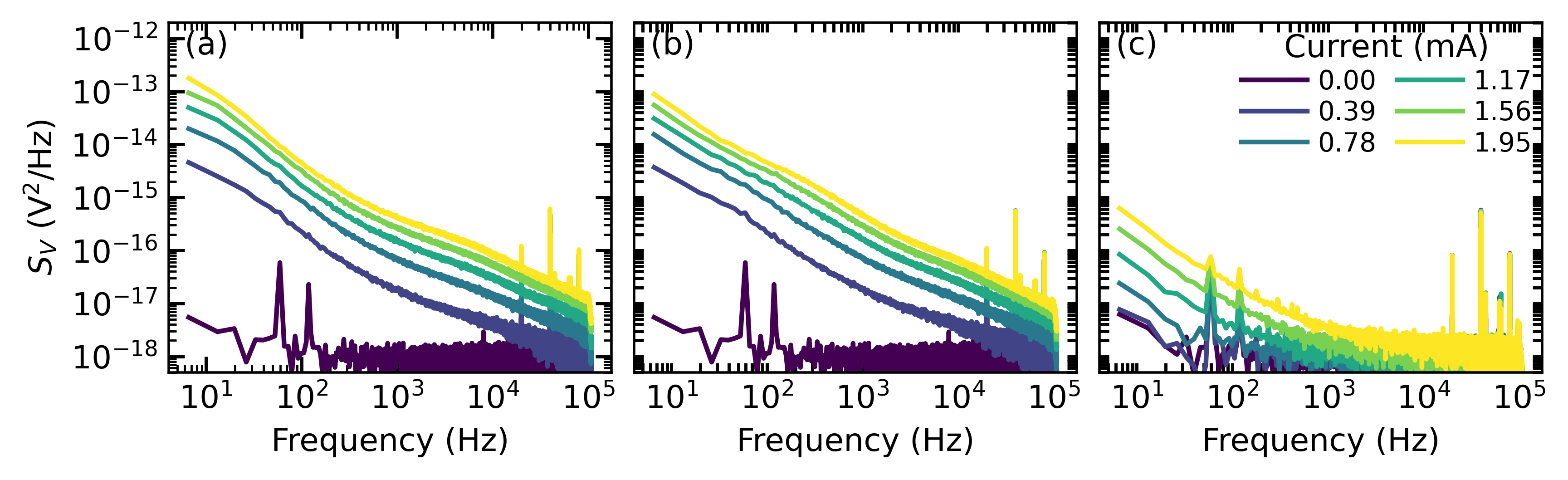}
    \caption{Raw noise measurements of the Nb junction with positive (a) and negative (b) dc, as well as ac (c) biases.}
    \label{fig:spectresRawDec}
\end{figure*}

\subsection{Average noise vs. bias}
\label{sec:NoiseVsBias}
Fig. \ref{fig:powers} shows the noise spectral densities averaged between $6.8$ and $30$ Hz, against the bias current for $S_\pm$ (a,b) and $S_{AC}$ (c). In agreement with Eqs.(\ref{eq:Sac},\ref{eq:S_symm}), $S_+\propto I_0^2$ and $S_{AC}\propto I_0^4$ for both junctions. $S_-\propto I_0^3$ is observed for the Nb junction but is almost absent for the Al junction. 
\begin{figure}
    \centering
    \includegraphics[width=\linewidth]{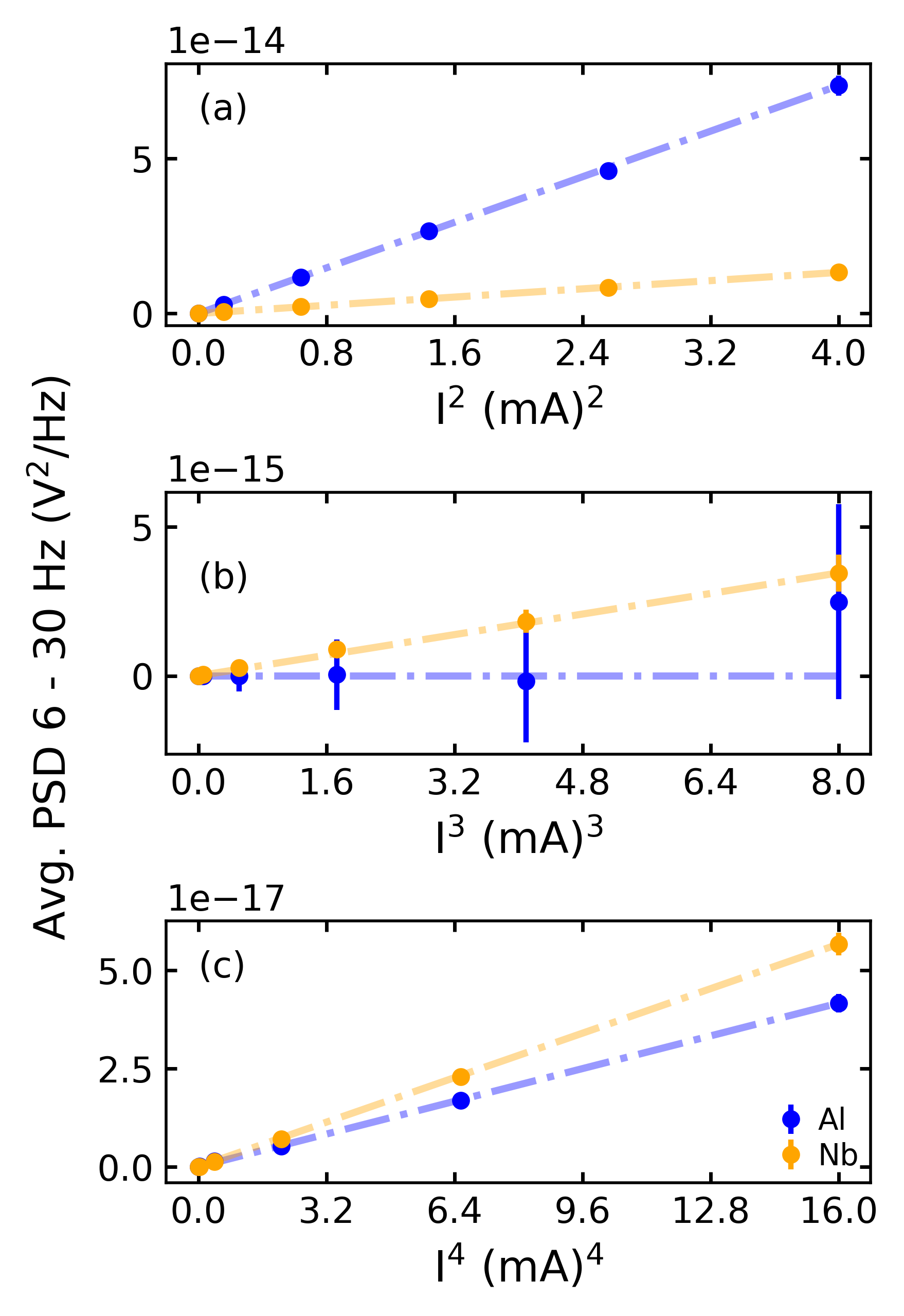}
    \caption{Average noise spectral density of the Al junction (blue) and Nb junction (orange). (a) $S_+$, (b) $S_-$, (c) $S_{AC}$}
    \label{fig:powers}
\end{figure}

\subsection{Extraction of $S_R^{0,0}$, $S_R^{1,1}$ and $S_R^{0,1}$}
\label{sec:extractSR}
From the measurements of $S_\pm$ and $S_{AC}$ we extract $S_R^{0,0}$, $S_R^{1,1}$ and $S_R^{0,1}$. $S_R^{0,0}$ and $S_R^{0,1}$ are obtained by taking $S_+$ and $S_-$, and limiting them to currents where the noise follows $I^2$ and $I^3$, before normalizing them by the current and other numerical factors. $S_R^{1,1}$ is obtained similarly, by normalizing $S_{AC}$, this time with no ceiling on current. For all spectra, only the points where data is greater than the uncertainty ($1\sigma$) are kept.

The resulting $S_R^{0,0}$, $S_R^{1,1}$ and $S_R^{0,1}$ are shown in Fig. \ref{fig:flutuatorsPSD}. Measured $S_R^{1,1}$ (dotted lines), i.e. the variances of $R_1$, appear nearly identical for both the Al (blue) and Nb (orange) junctions even though the average values $\langle R_1\rangle$ differ by more than an order of magnitude. Their measured levels are orders of magnitude greater than the other spectral densities. $S_R^{0,0}$ (solid line), i.e. the variances of $R_0$ are similar for both junctions although as seen previously, their slope varies. The results for $S_R^{0,1}$ (dashed lines), which measures the correlation between the fluctuations of $R_0$ and those of $R_1$ are very different: correlations are nonexistent for the Al junction and clearly measurable for the Nb junction. For the latter, it is even negative, which corresponds to anti-correlations,  in the gray zone from 90 Hz to 1300 Hz.
\begin{figure}
    \centering
    \includegraphics[width=\linewidth]{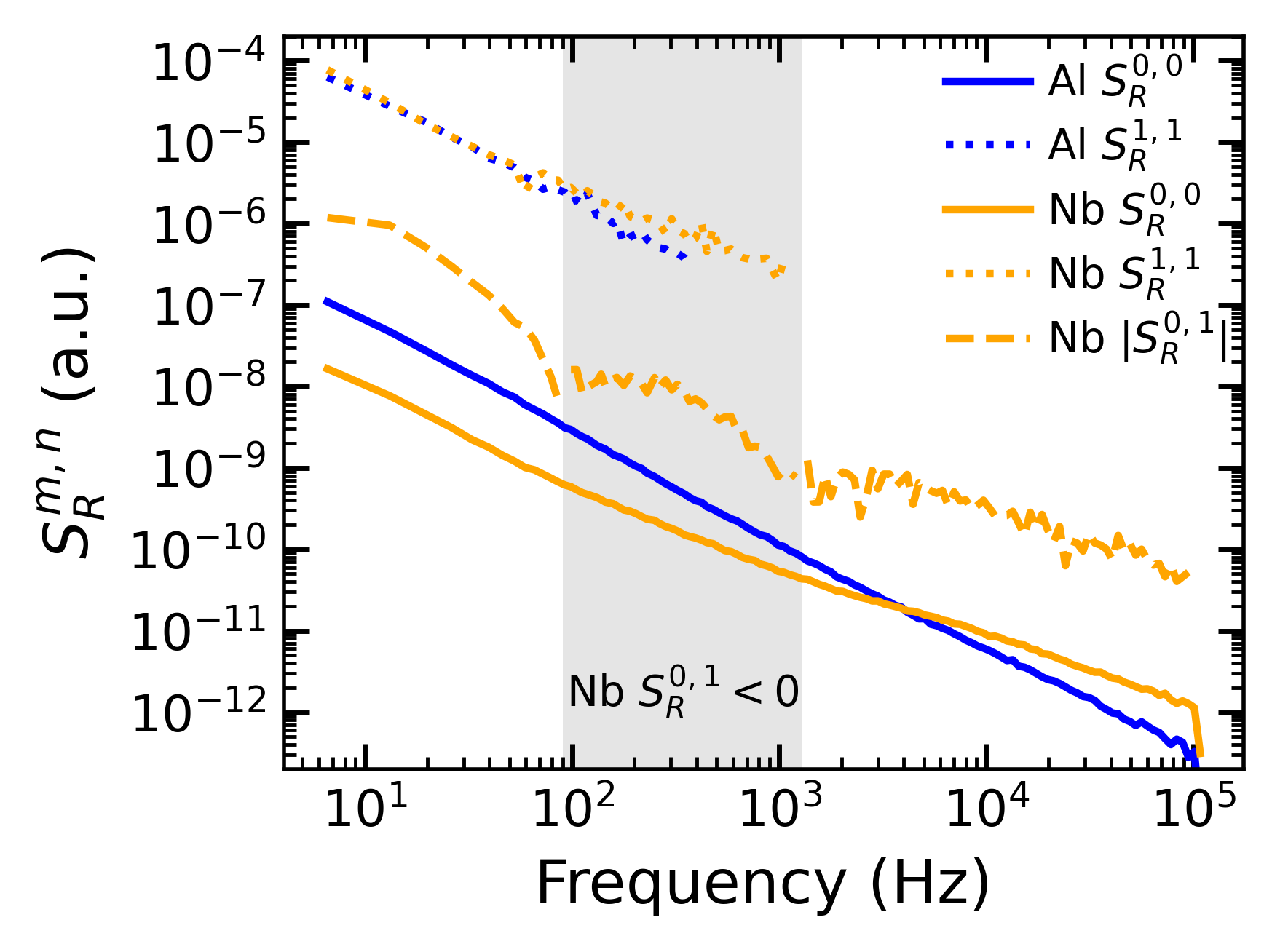}
    \caption{$S_R^{0,0}$ (solid lines), $S_R^{1,1}$ (dotted lines) and $S_R^{0,1}$ (dashed lines) for the Al (blue) and Nb (orange) junctions. $S_R^{0,1}$ takes negative values for the Nb junction in the shaded area. Data points were downsampled for visibility.}
    \label{fig:flutuatorsPSD}
\end{figure}

\subsection{Fluctuations of the nonlinear resistance coefficients in the BDR framework}
\label{sec:BDRframework}

From Eqs. (\ref{eq:VI},\ref{eq:brinkman}) we get,
\begin{equation}
R_0 = 4\frac{B}{A}\frac{d}{\sqrt{\phibar}}\exp(\eta)
\end{equation}
\begin{equation}
R_1 = \frac23\frac{B^2}{A}\frac{d^3\deltaphi}{\phibar^{5/2}}\exp(2\eta)
\end{equation}
\begin{equation}
R_2 = \frac{16}{6}\frac{B^3}{A}\frac{d^5}{\phibar^{5/2}}\left[\frac{1}{12}\frac{\deltaphi^2}{\phibar^{2}} - 1\right]\exp(3\eta)    
\end{equation}
with $A = \sqrt{2m_eq}/\hbar$, $\eta = 2Ad\sqrt{\phibar}$ and $B = \pi^2\hbar/q^2S$.

We can write the fluctuations of $R_{0,1}$ as
\begin{equation}
\delta R_0(t) = \left[ k_1\delta d(t) + k_2\delta\phibar(t) \right]\langle R_0\rangle \label{eq:deltaR0}
\end{equation}
\begin{equation}
\delta R_1(t) = \left[ k_3\delta d(t)+ k_4\delta\phibar(t) + \frac{\delta\deltaphi(t)}{\deltaphi} \right]\langle R_1\rangle \label{eq:deltaR1}    
\end{equation}
where $k_1 = \frac{1+\eta}{d}$, $k_2 = \frac{\eta-1}{2\phibar}$, $k_3 = \frac{3+2\eta}{d}$ and $k_4 = \frac{2\eta-5}{2\phibar}$.

And simply get the following spectral densities,
\begin{equation}
\frac{S_R^{0,0}(f)}{\langle R_0\rangle^2} = k_1^2 S_{dd} + k_2^2 S_{\phibar\phibar} \label{eq:SR00}    
\end{equation}
\begin{equation}
\frac{S_R^{1,1}(f)}{\langle R_1\rangle^2} = k_3^2 S_{dd} + k_4^2S_{\phibar\phibar} + \frac{S_{\deltaphi\deltaphi}}{\deltaphi^2} + \frac{2k_4}{\deltaphi} S_{\phibar\deltaphi} \label{eq:SR11}    
\end{equation}
\begin{equation}
\frac{S_R^{0,1}(f)}{\langle R_0\rangle \langle R_1\rangle} = k_1k_3 S_{dd} + k_2k_4 S_{\phibar\phibar} + \frac{k_2}{\deltaphi} S_{\phibar\deltaphi} . \label{eq:SR01}
\end{equation}

\subsection{Analysis with $S_{dd}\neq0$}
\label{sec:dnotzero}
Here we consider the effects of the fluctuations of the barrier thickness $d$, i.e. $S_{dd}\neq0$. Assuming that those fluctuations are uncorrelated with that of the barrier height $\phibar$ and asymmetry $\deltaphi$, the conclusions of the main text hold. For a symmetric junction ($\deltaphi\to 0$), only the last term of Eq. (\ref{eq:deltaR1}) remains since $\langle R_1\rangle \propto \deltaphi$. In that case, $S_{\deltaphi\deltaphi} \propto S_R^{1,1}(f)$ and $S_{\phibar\deltaphi} \propto S_R^{0,1}(f)$: the noise of a symmetric junction provides direct measurements of the spectral densities of the fluctuations of $\deltaphi$ and the correlations between the fluctuations of $\phibar$ and $\deltaphi$. The experiments on the Al junction demonstrate that even though $\deltaphi$ is almost zero on average, its fluctuations are important and not correlated with those of $\phibar$.

For the asymmetric junction, measurements show the presence of nonzero $S_R^{0,1}$. For part of the spectrum, $S_R^{0,1}$ is even negative. From Eq. (\ref{eq:SR01}), a negative $S_R^{0,1}$ is achievable in two ways. The first requires $\eta\in [1,2.5]$ since the product $k_2k_4$ becomes negative, but the fit of Eq. (\ref{eq:brinkman}) gives $\eta\approx12$. Even then, the change of sign of $S_R^{0,1}$ would not be explained. The second would be a nonzero $S_{\phibar\deltaphi}$, as we will demonstrate.

Around 10 Hz $S_R^{0,1}$ is up to 50 times greater than $S_R^{0,0}$. From 100 to 1000 Hz, $S_R^{0,1}$ is negative and close to 20 times the magnitude of $S_R^{0,0}$. Comparing Eqs. (\ref{eq:SR00}) and (\ref{eq:SR01}), it is clear that $S_{\phibar\deltaphi}$ has to be nonzero. The results of $S_R^{0,0}$ give upper bounds on $S_{\phibar\phibar}$. We can isolate $S_{dd}$ from Eq. (\ref{eq:SR00}) and insert it in Eq. (\ref{eq:SR01}). We get,
\begin{equation}
\frac{S_R^{0,1}}{R_0R_1} - \frac{k_2}{\deltaphi}S_{\phibar\deltaphi} = \frac{k_3}{k_1}\frac{S_R^{0,0}}{R_0^2} + \Gamma k_2^2 S_{\phibar\phibar}
\end{equation}
with $\Gamma = \frac{k_4}{k_2}-\frac{k_3}{k_1} = -\frac{4\eta+2}{\eta^2-1}$. By definition, since $S_{dd},S_{\phibar\phibar}\geq0$, Eq. (\ref{eq:SR00}) forces $k_2^2S_{\phibar\phibar}\leq S_R^{0,0}/R_0^2$. So, with $\eta > 1$,
\begin{equation}
\frac{k_4}{k_2}\frac{S_R^{0,0}}{R_0^2} \leq\ \frac{S_R^{0,1}}{R_0R_1} - \frac{k_2}{\deltaphi}S_{\phibar\deltaphi} \leq \frac{k_3}{k_1}\frac{S_R^{0,0}}{R_0^2}
\end{equation}
For $\eta\approx12$ we find $k_3/k_1\sim2.1$. Therefore $S_{dd}$ and $S_{\phibar\phibar}$ can be responsible for at most 5\% of the $S_R^{0,1}$ signal around 10 Hz. $S_{\phibar\deltaphi}$ appears in asymmetric junctions but not in symmetrical junctions. We observed these results for multiple junctions.


\subsection{Implications on barrier fluctuations}
\label{sec:implications}
Using the definitions of $\phibar$ and $\deltaphi$, we can get back to the fluctuations of the levels on each side of the barrier, $\varphi_1$ and $\varphi_2$. In this section we again assume no fluctuations of the barrier thickness $d$, i.e. $S_{dd}=0$. We get,
\begin{eqnarray}
    S_{\phibar\phibar} &=& \frac14(S_{\varphi_1\varphi_1}+S_{\varphi_2\varphi_2}+2S_{\varphi_1\varphi_2})  \label{eq:Spb12}\\
    S_{\deltaphi\deltaphi} &=& S_{\varphi_1\varphi_1}+S_{\varphi_2\varphi_2}-2S_{\varphi_1\varphi_2} \label{eq:Sdp12} \\
    S_{\phibar\deltaphi} &=& \frac12(S_{\varphi_1\varphi_1}-S_{\varphi_2\varphi_2}) \label{eq:Sdppb12}
\end{eqnarray}

From Eqs. (\ref{eq:Spb12},\ref{eq:Sdp12}), it is clear that $S_{\varphi_1\varphi_2}$ has to be negative in order to explain $S_{\deltaphi\deltaphi}$ being orders of magnitude above $S_{\phibar\phibar}$ in both junctions. Fig. \ref{fig:spectresPhis} shows the spectra of $S_{\varphi_1\varphi_1}$ (solid lines), $S_{\varphi_2\varphi_2}$ (dashed lines) and $S_{\varphi_1\varphi_2}$ (dotted lines) for the Al (blue) and Nb (orange) junctions. We find $S_{\varphi_1\varphi_1} = S_{\varphi_2\varphi_2}$ for the Al junction, which is expected since $S_{\phibar\deltaphi} = 0$. More importantly $S_{\varphi_1\varphi_2}$ is negative for the entirety of the measured bandwidth in both junctions, meaning the fluctuations of $\varphi_1$ and $\varphi_2$ are anti-correlated.
\begin{figure}
    \centering
    \includegraphics[width=\linewidth]{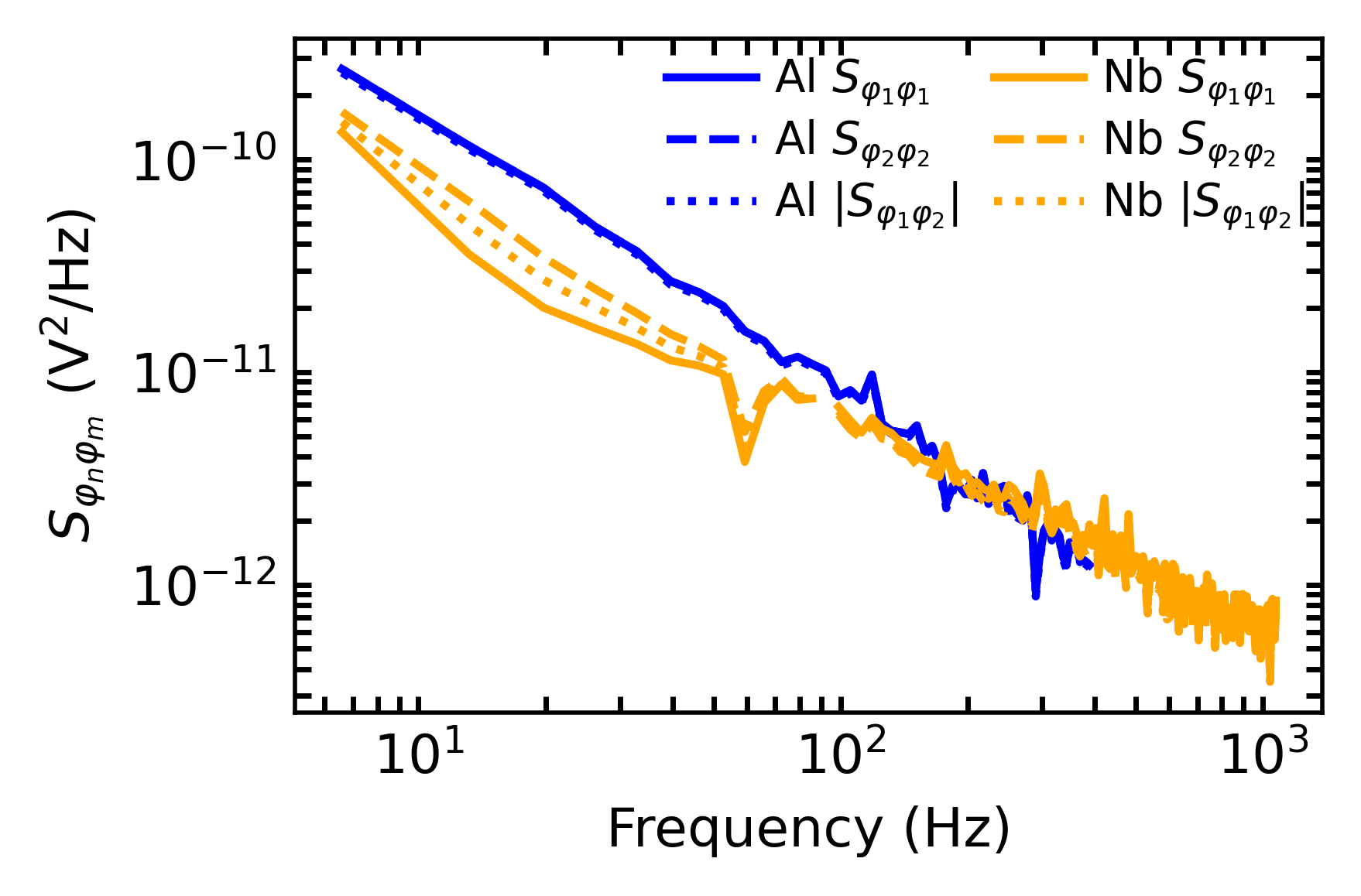}
    \caption{Spectra of $S_{\varphi_1\varphi_1}$ (solid lines), $S_{\varphi_2\varphi_2}$ (dashed lines) and $S_{\varphi_1\varphi_2}$ (dotted lines) for the Al (blue) and Nb (orange) junctions.}
    \label{fig:spectresPhis}
\end{figure}
The correlation coefficient $\sigma_{\varphi_1\varphi_2} = S_{\varphi_1\varphi_2}/\sqrt{S_{\varphi_1\varphi_1}S_{\varphi_2\varphi_2}}$ of the fluctuations of $\varphi_1$ and $\varphi_2$ is presented in Fig.\ref{fig:correlationsPhis}. For both junctions, we find values extremely close to $-1$, meaning that the fluctuations of $\varphi_1$ and $\varphi_2$ are almost perfectly anti-correlated.
\begin{figure}
    \centering
    \includegraphics[width=\linewidth]{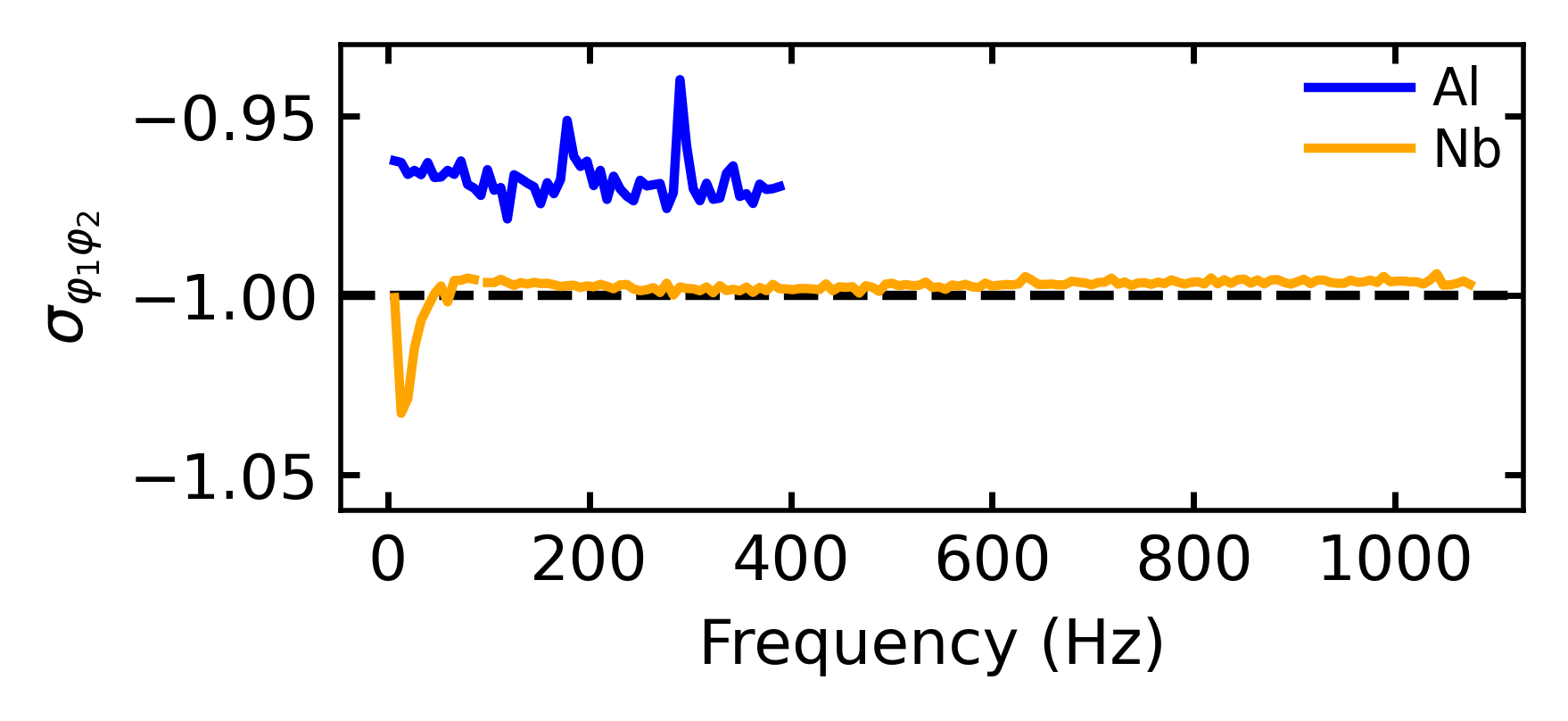}
    \caption{Correlation coefficient $\sigma_{\varphi_1\varphi_2}$ of the fluctuations of $\varphi_1$ and $\varphi_2$ for the Al (blue) and Nb (orange) junctions.}
    \label{fig:correlationsPhis}
\end{figure}

\end{document}